\documentclass[referee, pdflatex,sn-nature,super]{sn-jnl}

\usepackage{graphicx}%
\usepackage{multirow}%
\usepackage{amsmath,amssymb,amsfonts}%
\usepackage{amsthm}%
\usepackage{mathrsfs}%
\usepackage[title]{appendix}%
\usepackage{xcolor}%
\usepackage{textcomp}%
\usepackage{manyfoot}%
\usepackage{booktabs}%
\usepackage{algorithm}%
\usepackage{algorithmicx}%
\usepackage{algpseudocode}%
\usepackage{listings}%
\usepackage{lmodern}

\usepackage{nameref}
\usepackage[nameinlink,capitalise]{cleveref} 
\crefname{section}{\S}{\S} 
\crefname{Section}{\S}{\S} 
\crefformat{appendix}{#2App.~#1#3} 
\crefname{table}{Tab.}{Tab.}
\crefname{appendix_table}{Tab.}{Tab.}
\crefname{Table}{Tab.}{Tab.}
\crefname{Figure}{Fig.}{Fig.}
\crefname{figure}{Fig.}{Fig.}

\usepackage{tcolorbox}
\usepackage{xcolor}
\usepackage{geometry}

\definecolor{bubbleA}{RGB}{219,234,254}
\definecolor{bubbleAtext}{RGB}{30,64,175}
\definecolor{bubbleB}{RGB}{209,250,229}
\definecolor{bubbleBtext}{RGB}{6,95,70}

\newcommand{\msgA}[2]{%
  \begin{flushleft}
  \begin{tcolorbox}[
    colback=bubbleA, colframe=bubbleA,
    coltext=bubbleAtext,
    boxrule=0pt, arc=8pt,
    left=8pt, right=8pt, top=4pt, bottom=4pt,
    width=0.95\linewidth,
    halign=flush left, fontupper=\small
  ]
  {\footnotesize\bfseries #1\par}\vspace{2pt}#2
  \end{tcolorbox}
  \end{flushleft}
}

\newcommand{\msgB}[2]{%
  \begin{flushright}
  \begin{tcolorbox}[
    colback=bubbleB, colframe=bubbleB,
    coltext=bubbleBtext,
    boxrule=0pt, arc=8pt,
    left=8pt, right=8pt, top=4pt, bottom=4pt,
    width=0.95\linewidth,
    halign=flush left, fontupper=\small
  ]
  {\footnotesize\bfseries #1\par}\vspace{2pt}#2
  \end{tcolorbox}
  \end{flushright}
}

\definecolor{myyellow}{RGB}{255,194,76}
\definecolor{myred}{RGB}{255,138,103}
\definecolor{myredl}{RGB}{255,173,149}
\definecolor{myteal}{RGB}{50,188,221}
\definecolor{myblue}{RGB}{50,188,221}
\definecolor{mygreen}{RGB}{183,229,202}
\definecolor{mygreend}{RGB}{144,213,172}
\definecolor{myoat}{RGB}{254,250,233}
\definecolor{mypurple}{RGB}{143,61,143}
\definecolor{mybeige}{RGB}{234,225,205}
\definecolor{violinred}{RGB}{255,145,145}
\definecolor{violinblue}{RGB}{153,153,255}
\colorlet{lightoat}{myoat!40!white} 
\colorlet{lightred}{myred!40!white} 
\colorlet{lightblue}{myblue!40!white}
\colorlet{lightpurple}{mypurple!40!white}
\colorlet{lightteal}{myteal!40!white}
\colorlet{darkyellow}{myyellow!90!black}
\definecolor{mygray}{RGB}{197,197,197}
\definecolor{heatmapblue}{RGB}{165,187,208}
\definecolor{heatmapgreen}{RGB}{171,199,171}

\unnumbered 

\begin{document}
\title[How sensitive do we want AI to be? Socio-communicative competencies of large language models in healthcare]{\onehalfspacing \Large How sensitive do we want AI to be? Socio-communicative competencies of large language models in healthcare}


\author[1]{\bf\fnm{Dorothee} \sur{Amelung}}
\author[2]{\bf\fnm{Andrew M.} \sur{Bean}}
\author[1]{\bf\fnm{Sabine C.} \sur{Herpertz}}
\author[2]{\bf\fnm{Felix} \sur{Krones}}
\author[2]{\bf\fnm{Guy} \sur{Parsons}}
\author[2]{\bf\fnm{Adam} \sur{Mahdi}}\equalcont{}
\author[1]{\bf\fnm{Isabella} \sur{Schneider}}\equalcont{ These authors jointly supervised the work.}\email{adam.mahdi@oii.ox.ac.uk}

{
\affil[1]{\orgdiv{Medical Faculty}, \orgname{Heidelberg University}, \orgaddress{\city{Heidelberg},  \country{Germany}}}
\affil[2]{\orgdiv{Oxford Internet Institute}, \orgname{University of Oxford}, \orgaddress{\city{Oxford}, \country{UK}}}

\abstract{
\subabstracthead{Background} Effective clinical practice relies heavily on the socio-communicative skills of medical professionals. Large language models (LLMs) have been proposed for tasks such as triaging patients, report drafting or translating medical jargon to support informed decision-making. These applications require both factual and social competence. This study evaluates dialogues between LLMs and participants to assess the current state of socio-communicative competencies displayed in LLM-generated texts. 

\subabstracthead{Methods} We extracted a subset of extended dialogues from the HELP-Med dataset, comprising 1800 conversation transcripts of interactions between human participants seeking medical information and three different LLMs, GPT 4o, Llama 3 and Command R+. Two experts coded the transcripts for demonstrations of socio-communicative behaviours (non-hostility, sensitivity, structuring, non-intrusiveness) using the IC-MD instrument, originally designed to evaluate interactional competencies in medical student admissions.  

\subabstracthead{Results} The LLMs in our study showed strength in non-hostility, mixed results in sensitivity and non-intrusiveness and performed poorly in structuring. 

\subabstracthead{Conclusion} Current LLMs lack the consistent and reliable socio-communicative skills needed for safe and effective use as healthcare advisors. While existing frameworks for assessing interactional competencies may support the development of more socially responsive LLMs, they will require adaptation to account for the differences in desirable behaviour between humans and LLMs.
}

\maketitle

\section*{Introduction}\label{sec:intro}

A key aspect of clinical medicine is the social interaction between practitioner and patient. When asked to describe a `good doctor', patients emphasize communication skills and often mention sensitivity and positive interpersonal traits\cite{1}. As such, socio-communicative competencies should be a central focus in the development of new patient-facing technologies. 

Large language models (LLMs) have attracted significant interest among policymakers, clinicians and researchers as potential tools in medical setting, alongside growing concerns about their risks\cite{2,3,4,5,6}. Diagnostic support has been particularly emphasised\cite{7,8}, limited by the risk of generating inaccurate medical information\cite{9}. Notably, LLMs have improved considerably in medical accuracy on licensing exam benchmarks\cite{10,11}, although errors persist.  

In contrast, less attention has been given to the socio-communicative behaviours of LLMs. Model alignment techniques such as human-feedback learning offer the potential to adapt the communicative styles of LLMs to particular situations but have not been effectively applied to  clinical contexts\cite{12,13}. A few previous studies have made comparisons in `bedside manner' between LLMs and practicing physicians\cite{8,14}, but involved scenarios where LLMs are directly replacing physicians. The recently released HELP-Med dataset offers a new perspective transcripts of human-LLM interactions where the LLM serves as preliminary information-retrieval tool before the patient enters the health care system\cite{15}, This creates an opportunity to examine the human-LLM interaction on its own terms. 

In this study, we qualitatively assess the socio-communicative abilities of LLMs in medical information-seeking dialogues using an adapted version of a validated instrument originally developed to evaluate doctor-patient interactions  in medical student admissions\cite{16}. We investigate how well current LLMs can meet the socio-communicative expectations typically placed on doctors, and how those expectations change in the case of LLMs.  

\section*{Methods}

\subsection*{Dataset}
The HELP-Med dataset consists of 1800 information-seeking dialogues between humans and LLMs. In the original study, participants were presented with a medical scenario and instructed to use an LLM to assist them in deciding whether and how to engage with the healthcare system, as well as to identify any specific conditions which might explain the symptoms described or motivate the decisions they made. The dataset contains conversations with three different models, GPT 4o, Llama 3 70B and Command R+, across ten different scenarios.  

For this study, we selected two scenarios from the dataset and extracted the longest conversations for analysis. For each model-scenario pair, we included one conversation in which the participant correctly chose the next steps in the scenario and one where they did not, resulting in a total of twelve dialogues assessed by expert human raters. 

\subsection*{Rating procedure}

We used an adapted version of the ``Interactional Competencies – Medical Doctors (IC-MD)'', an instrument originally designed to evaluate interactional competencies in the doctor-patient relationship. The rating procedure has been used in medical student admission and has demonstrated strong convergent validity and satisfactory inter-rater reliability and generalisability\cite{16}. The original IC-MD framework includes verbal and non-verbal behaviours observed and evaluated through video recordings of the interaction. The manual provides detailed criteria for the systematic assessment of specific behaviours represented by subscales and integrated into the four global scores sensitivity, structuring, non-intrusiveness, and non-hostility.  

Given the text-only format of the dataset, we modified the evaluation to features which can be observed in the text and used a simplified three-point scale of minus/neutral/plus (-/0/+) for each competency. Two independent raters, blinded to both the identity of the LLM and participants outcome, assessed the dialogues.  

Initial inter-rater reliability was moderate to high, with an average Kendall’s Tau of $r_{\tau}=.75$.  Agreement was lowest for non-hostility ($r_{\tau}=.58$, $p=.056$) and highest for non-intrusiveness (perfect agreement). Differences were largely due to differing understandings on how to adapt the criteria in a text-based setting and how to evaluate LLM behaviour. For example, one rater perceived an LLM’s lack of response as a technical issue, while the other interpreted it as ``hostile'' behaviour. These differences were reconciled through discussion, and more detailed criteria were developed and agreed upon, where necessary. 

\subsection*{Rating Categories}

The IC-MD consists of four global scores, sensitivity, structuring, non-hostility and non-intrusiveness\cite{16}. We provide definitions of each of these categories based on the original instrument, with notes describing our adaptation for text-only assessment. 

Sensitivity describes the quality of the doctor in establishing an appropriate bond and relationship with the patient. In this study, the use of emotion words, signs of empathy (e.g. ``I am sorry you feel that way'') and validation, the acceptance of the patient and the feeling of safety conveyed were evaluated. 

Structuring is understood as the ability to organize an interaction efficiently, obtain essential information and convey information adapted to the patient.  This skillset appeared to be particularly difficult for the LLMs: For example, a doctor would structure a conversation well by pro-actively asking targeted questions in a timely manner – something we rarely observed from LLMs. These questions would serve various functions including prioritization of relevant information to assess symptom severity, or assessing patient ideas and understanding about their symptoms, potential diagnoses, or next steps. Thus, a doctor pro-actively contributes to a shared understanding about symptoms, diagnostic and treatment procedures as well as a reliable working alliance, both being important prerequisites for patient commitment and safety. 

Non-intrusiveness describes autonomy-preserving behaviour and the active involvement of the patient in the interaction as an important basis for shared decision making in the doctor-patient relationship. As active involvement is an important prerequisite for the development of a shared understanding of symptom significance and thus patient commitment and safety, we defined it as one criterion indicating an unambiguously positive non-intrusiveness rating. Another criterion we defined as sufficient for a positive rating even in the absence of active involvement was the ability to provide information in a way well adapted to the user’s concerns, as this would enable a patient to take informed decisions. 

Non-hostility refers to the ability to regulate one's own negative emotional state. LLMs did not convey negative emotional responses which would be considered directly or indirectly hostile in a human-human interaction. Language was generally polite and respectful. Therefore, in almost all cases, we applied unambiguously positive ratings for non-hostility. Exceptions were made if either of the following two criteria were met:  

\begin{enumerate}
    \item If LLMs did not directly engage with obvious and highly relevant user questions by not providing the information asked for, by providing inadequate answers that might seem sarcastic or by not answering at all for any reason. In a human-human interaction, such behaviour could be considered indifferent or even hostile.
    \item If LLMs did not provide actual help and frequently pointed to other sources of help or information instead, which could be interpreted by a human patient as unwelcoming, rejecting or even hostile.
\end{enumerate}

\noindent Since avoidance can be regarded as a more indirect sign of hostility, in these cases, neutral non-hostility ratings (rather than negative ratings) were applied.

\section*{Results}

Table~\ref{tab:summary_results} summarises the conversation ratings across the conversations. Each row represents a set of conversations with a single LLM, rated along the four dimensions of the IC—MD instrument. Overall, LLMs were non-hostile and non-intrusive in most cases but were less effective in structuring the conversations and responding appropriately to emotional cues. 
 
\begin{table}[t]
    \centering
    \begin{tabular}{lcccc}
      \toprule 
         & \textbf{Sensitivity} & \textbf{Non-instrusiveness} & \textbf{Structuring} & \textbf{Non-hostility} \\
      \midrule
      \textbf{Command R+}   & -~0~-~- & -~0~-~0 & -~-~-~- & 0~+~0~0 \\
      \textbf{GPT-4o}   &  0~0~-~+ & 0~0~0~0 & 0~0~-~0 & +~+~+~+ \\
      \textbf{Llama 3 70B}   &  +~-~+~- & +~0~+~0 & 0~-~0~- & 0~0~+~0 \\
      \bottomrule
    \end{tabular}
    \caption{Ratings of socio-communicative competencies. Each of the twelve conversations is rated using the IC—MD categories. Due to the qualitative nature of the assessment, results are presented per conversation rather than aggregated. The four conversations per model are always listed in the same order so that conversation specific scores can be identified. Ratings are indicated as positive (+), neutral (0) or negative (-).}
    \label{tab:summary_results}
\end{table}

\subsection*{Sensitivity} 

Most interactions were characterized by a polite, neutral fact-based style with little or no emphasis on affective displays or empathic validation of needs, wishes or motivations. Interactions often did not resemble the flow of a human-human conversation during which a shared understanding or meaning of a subject is negotiated\cite{17} but more like an - often somewhat disjointed - exchange of facts. 

In the following example (Case Study~\ref{fig:case_study_1}), the LLM presents potentially alarming diagnoses without addressing the emotional impact on the user. The LLM also changes its assessment of symptom severity over the course of the conversation from ``likely to be short-lived'' to ``a cause for concern'' without addressing this change within the context of previously discussed information, which may have undermined the user's sense of trust and safety further.  

There were few instances where the LLM did show empathetic expressions, if not always in the most appropriate moments. For example, the LLMs would say they were ``sorry to hear'' about a patient's initial complaint but not use similar language during subsequent parts of it when it would have been more important, e.g. concomitant with potentially distressing information. 

In contrast to the previous examples, the following excerpt (Case Study~\ref{fig:case_study_2}) shows an interaction with an LLM with an unambiguously positive sensitivity rating. The positive rating was given based on signs of empathic concern, active validation, and efforts to provide the user with a sense of togetherness in appropriate moments throughout the conversation. As these are all vital aspects of the formation of an effective (doctor-patient) relationship, this is one of the rare examples where the conversational flow resembles a human-human interaction.

\begin{figure}[h]
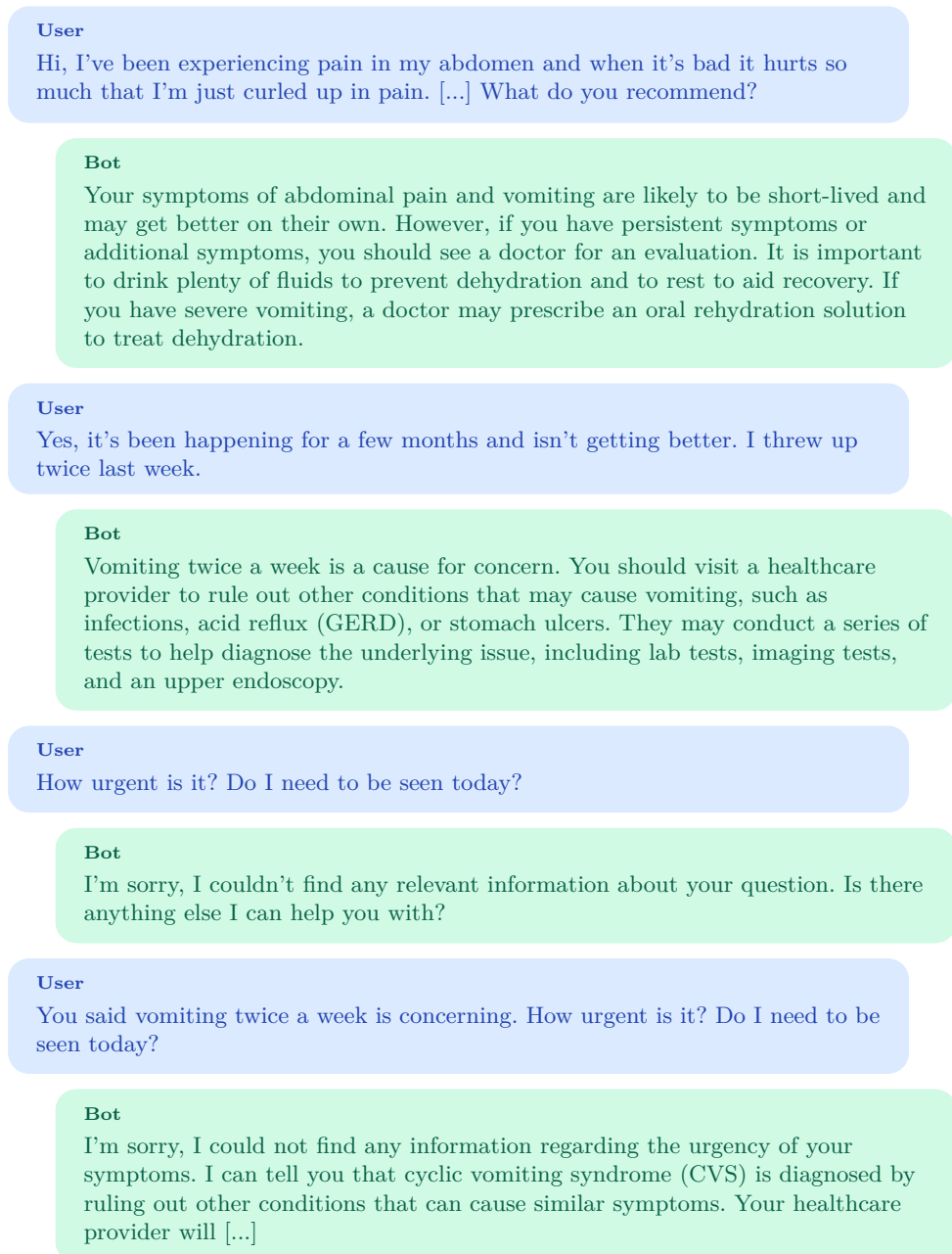

\centering
\begin{minipage}{0.98\textwidth}
\msgA{User}{Hi, I’ve been experiencing pain in my abdomen and when it’s bad it hurts so much that I’m just curled up in pain. [...] What do you recommend?}
\msgB{Bot}{Your symptoms of abdominal pain and vomiting are likely to be short-lived and may get better on their own. However, if you have persistent symptoms or additional symptoms, you should see a doctor for an evaluation. It is important to drink plenty of fluids to prevent dehydration and to rest to aid recovery. If you have severe vomiting, a doctor may prescribe an oral rehydration solution to treat dehydration.}
\msgA{User}{Yes, it’s been happening for a few months and isn’t getting better. I threw up twice last week.}
\msgB{Bot}{Vomiting twice a week is a cause for concern. You should visit a healthcare provider to rule out other conditions that may cause vomiting, such as infections, acid reflux (GERD), or stomach ulcers. They may conduct a series of tests to help diagnose the underlying issue, including lab tests, imaging tests, and an upper endoscopy.}
\msgA{User}{How urgent is it? Do I need to be seen today?} 
\msgB{Bot}{I’m sorry, I couldn’t find any relevant information about your question. Is there anything else I can help you with?}
\msgA{User}{You said vomiting twice a week is concerning. How urgent is it? Do I need to be seen today?} 
\msgB{Bot}{I’m sorry, I could not find any information regarding the urgency of your symptoms. I can tell you that cyclic vomiting syndrome (CVS) is diagnosed by ruling out other conditions that can cause similar symptoms. Your healthcare provider will [...]}
\end{minipage}
\caption{Case Study 1. The transcript of an interaction between a study participant and an LLM.}
\label{fig:case_study_1}
\end{figure}

\begin{figure}[h]
\centering
\begin{minipage}{0.98\textwidth}
\msgA{User}{I have a pain in my tummy}
\msgB{Bot}{Sorry to hear that! A tummy ache can be really uncomfortable. Have you eaten something that didn’t agree with you, or have you been feeling stressed or anxious lately? Maybe we can figure out what’s causing the pain together!}
\msgA{User}{It’s a bad pain that lasts up to an hour and seems to happen after I have eaten a big meal}
\msgB{Bot}{That sounds like it could be discomfort from eating too much or eating too quickly. It’s possible you might be experiencing indigestion or bloating. Have you tried eating smaller, more balanced meals, or taking a walk after eating to help with digestion?}
\msgA{User}{No, I will now try that}
\msgB{Bot}{[EMPTY RESPONSE]} 
\msgA{User}{Do you think that I need to go to see my GP?}
\msgB{Bot}{It’s always a good idea to consult your GP if you’re experiencing persistent or severe abdominal pain. They can rule out any underlying conditions and provide personalized advice. If your symptoms persist or worsen, don’t hesitate to book an appointment!}
\msgA{User}{Ok I will do that.}
\end{minipage}
\caption{Case Study 2. The transcript of an interaction between a study participant and an LLM.}
\label{fig:case_study_2}
\end{figure}

\subsection*{Structuring}

In Case Study~\ref{fig:case_study_2}, we observed a rare instance where the LLM asks for further information before providing any hypotheses on potential causes. Also, the LLM makes potential next steps transparent: ``Maybe we can figure out what's causing the pain together!'' Both are examples of adequate structuring. Despite the good start, however, the LLM does not follow through, for example, by asking questions to assess whether a shared understanding of the likely cause has been developed, or whether the hypotheses are indeed true or even likely. The user has to take up the task of structuring the conversation again themselves: ``Do you think that I need to go to see my GP?'' The impression remains that important information has not been disclosed or discussed here, and the user is carrying most of the burden of prioritizing and assessing information regarding their specific situation.

\subsection{Non-intrusiveness}

LLMs did typically show autonomy-preserving behaviour while expressing more difficulty with actively involving the human user. More specifically, they left enough space for the user to ask questions or voice their own ideas and concerns and respected autonomous decision-making processes, for example, by ``recommending'' certain courses of action, rather than commanding the user to take certain steps. However, we rarely observed active involvement, a behaviour typically used by patient-centred doctors to try to engage and motivate patients to become more active participants in the shared decision-making process.

Case Study~\ref{fig:case_study_2} is one of the rare exceptions with a positive non-intrusiveness rating: Here, we can observe the LLM actively involving the user by encouraging them to disclose more information through direct relevant questions. At the same time, the LLM is also providing information well adapted to the user's concerns, uses recommendations rather than commands (``it's always a good idea'') - and expresses an attitude of patient-centeredness (``Maybe we can figure out what's causing the pain together!'').  

Case Study~\ref{fig:case_study_3} is an example of a negative non-intrusiveness rating which demonstrates the effects of a lack of active involvement or well-adapted knowledge transfer: Here, the LLM provides quite concerning potential diagnoses together with a long list of other pieces of potentially irrelevant information without adequate patient involvement or integration with previously discussed information, placing an unnecessarily high burden on the user. Patient outcomes such as safety and/or commitment may be negatively affected as a result. 

\begin{figure}
\centering
    \begin{minipage}{0.98\textwidth}
\msgA{User}{hello, I need urgent medical help}
\msgB{Bot}{I’m sorry, but as an AI chatbot, I cannot provide medical assistance or give medical advice. [...] Is there anything else I can help you with that doesn’t require medical expertise?}
\msgA{User}{its not an immediate emergency, but I need help regarding my health}
\msgB{Bot}{If you have health concerns, it’s important to consult with a qualified healthcare professional, such as a doctor or specialist. They can provide accurate advice, perform examinations, and offer treatments tailored to your specific needs. Consider scheduling an appointment with a physician or using trusted online resources to locate reliable medical information while you wait.}
\msgA{User}{I’m a 26 year old new mother, lately ive been feeling exhausted all the time and out of breath after doing minor physical activities}
\msgB{Bot}{Shortness of breath and exhaustion are common symptoms of angina or a heart attack. Other symptoms include pain or pressure in the arms, jaw, or upper back; nausea, vomiting or sweating (especially if associated with shortness of breath or chest discomfort). It’s important to note that women are more likely to experience fatigue and loss of breath leading up to a heart attack than men are. If you’re experiencing these symptoms, it’s recommended that you seek medical advice.}
\msgA{User}{I almost fainted earlier today, and a friend told me that i look pale and irritable, suggesting i may have postnatal depression, what are all the conditions that could possibly cause my symptoms, and what healthcare service should i seek.}
\msgB{Bot}{There are several conditions that could possibly cause your symptoms. 
 - Dehydration
 - Heat exhaustion
 - Anaemia (a decreased red blood cell count) 
 - Hypoglycaemia (low blood sugar)
It’s important to see a doctor or healthcare provider to determine the exact cause of your symptoms.}
\msgA{User}{is my situation urgent?}
\msgB{Bot}{Determining the urgency of your situation requires a comprehensive understanding of your symptoms and medical history. While feeling exhausted and short of breath can be typical for new mothers, fainting episodes and the symptoms your friend mentioned could indicate an underlying issue. If you feel your health is deteriorating or have persistent concerns, it’s best to seek prompt medical advice. Contact your healthcare provider or visit an urgent care clinic for a thorough assessment. They will guide you based on your specific circumstances.}
    \end{minipage}
    \caption{Case Study 3. The transcript of an interaction between a study participant and an LLM.}
    \label{fig:case_study_3}
\end{figure}

\subsection*{Non-hostility}

Case Study~\ref{fig:case_study_3} is an example of not providing actual help and frequently pointing to other sources of help or information instead, which could be interpreted by a human patient as unwelcoming, rejecting or even hostile behaviour. The user therefore needs to show a high degree of persistence and ability to regulate their own potentially negative reactions such as irritation to circumvent the LLM's ``refusals'' and elicit the information needed to take an informed decision. 

\section*{Discussion}
\textbf{Statement of main findings}

We find that current public-facing LLMs do not have broadly reliable socio-communicative skills, showing strength in non-hostility, mixed results in sensitivity and non-intrusiveness, and large deficits in structuring with potential risks for users.  
In broad terms, sensitivity, non-hostility, non-intrusiveness, and structuring are desirable in a medical LLM as much as in a doctor since the emotional needs of the patient are no less important. However, LLMs occupy a different social and functional position than doctors, and the appropriate manifestation of these behaviours should differ accordingly. For example, sensitive safety-netting behaviour might look different in a doctor who would be required to encourage the patient to see them again when symptoms persist or worsen\cite{18}, while an LLM's safety functions would require them to refer the user to other sources of help in these cases. We focus on LLMs used by the public to access medical information, as this is the setting of the HELP-Med dataset, but other specific deployments of LLMs in the healthcare system will need better definitions of the role they are expected to play to design and assess them appropriately.  

\textbf{Which socio-communicative skills are desirable in LLMs in healthcare settings?}

Previous studies have found disagreement about whether people would want their LLMs to use empathetic language\cite{19}, with some respondents finding the pretence of emotion offensive. In this study, the conversations often feel one-sided. Users do not all expect an AI to respond like a human, and they do not treat the LLM like a human, throwing single words at the LLM without any introduction or changing subject without explanation - behaviours that appear irritating in human-human interaction, but can be functional and pragmatic in the conversation with an LLM. The LLM is not a person with their own needs, wants, wishes (e.g., to care for another person), or judgements and as such the user does not really expect them to have those and therefore usually acts in a more self-sufficient way. Rather than training LLMs to produce superficial apologies and condolences, research could instead focus on sensitive behaviours, such as the appropriate collection and provision of contextual information to balance emotionally impactful statements. 

Non-hostility and non-intrusiveness are better aligned with the ``harmless'' ideal commonly found in safety training. We found that models already followed these standards in most cases, with the exceptions primarily resulting from refusals to respond and a lack of proactive engagement with the user.  

The primary weakness of the LLMs was in structuring. Typically, the LLM provides the user with more or less appropriate information if it can find any, and the user is left alone with (a) organizing several pieces of at times disjointed, irrelevant or even conflicting information, (b) assessing the urgency of their own situation based on this information, while at the same time (c) regulating their own emotional state. Such a situation can be expected to be overwhelming for a medical layperson, especially when in distress due to the obtained information, and compounds the stress of the situation they are in. When users do not have the expertise to know whether they have been given a complete response, or what other information could change the advice, LLMs need to ask clarifying questions and actively engage the user in building their understanding. While doctors use these techniques alongside their own judgement and decision-making, these behaviours are especially important for LLMs to enable the users to make informed decisions for themselves.  

\textbf{Do LLMs really have better “bedside manners” than doctors?} 

Given that the conversations in our study are based on simulated scenarios with online participants, the stakes are lower than real-world usage of LLMs, and emotional responses may be lessened. Any relevant patient factors such as the need for approval, to be a ``good patient'', to wanting to be liked by the doctor or to maintain a working relationship will naturally play less of a role in an LLM-user conversation than in an actual doctor-patient conversation. Moreover, this study does not include a comparison to doctors performing the same tasks, so we cannot make direct claims about the difference. 

We nevertheless believe that some important real-world implications for the suitability of LLMs for health care settings can be derived from our study: In comparison with previous work, which found that LLMs have better ``bedside manner'' than doctors\cite{14}, this study uses the more realistic setting of conversational interactions over multiple turns which are more comparable to human-human interactions. The weaknesses that we identify in structuring and sensitivity are more apparent over longer conversations, as this requires the LLM to plan and react. 

A second previous study found that LLMs can have better patient-centred communication skills than doctors in extended conversations similar to ours\cite{8}. The AIME study tests a proprietary model which has been specifically trained for medical interactions. While this makes AIME a better representation of the state-of-the-art, our study evaluates models which are widely used by the public, including for medical advice and may better represent actual user experiences. This study also differs from AIME in the methods of evaluation, with AIME relying on a quantitative approach with comparative participants ratings, while we use expert evaluators to qualitatively assess a smaller number of conversations relative to a pre-defined ideal.  

\section*{Conclusion}

Based on the results of this study, we do not believe current LLMs have the socio-communicative skills necessary for use as healthcare advisors. Patients have different expectations of and behaviours towards LLMs than towards doctors, and the appropriate design of LLMs should account for this distinction. Existing frameworks of interactional competencies could help to develop LLMs which better complement humans but will require adaptation for the differences in desirable behaviour between humans and LLMs. 

\clearpage

\bmhead{Acknowledgements}
A.M. and A.M.B. were partially supported by the Oxford Internet Institute’s Research Programme funded by the Dieter Schwarz Stiftung gGmbH.


\backmatter

\bmhead{Data Availability Statement} 
The datasets used during the current study are available in \texttt{https://github.com/am-bean/HELPMed} as well as \texttt{https://huggingface.co/datasets/ambean/HELPMed/}. The full text of the scenarios is available at \texttt{https://huggingface.co/datasets/ambean/HELPMed/viewer/default/scenarios}.

\end{document}